\documentclass[12pt]{article}
\usepackage{graphicx}
\usepackage{cite}
\usepackage{epsfig}
\usepackage{amsfonts}
 \usepackage{amssymb}
\usepackage{amsmath}
\usepackage{slashed}

\usepackage{hyperref}
\usepackage{graphicx}       % Include figure files
\usepackage{dcolumn}        % Align table columns on decimal point
\usepackage{bm}            		 % bold math
\usepackage{amssymb}
\usepackage{amstext}
\usepackage{tensor}
\usepackage{color} 
\usepackage{transparent}

\date{\today}

\newcommand{\insertplot}[5]{\begin{figure}
 \hfill\hbox to 0.05in{\vbox to #5in{\vfill
 \inputplot{#1}{#4}{#5}}\hfill}
 \hfill\vspace{-.1in}
 \caption{#2}\label{#3}
 \end{figure}}
 \newcommand{\inputplot}[3]{
 \special{ps: plotfile #1}
}
\newcounter{fig}

\newcommand{\beq}{\begin{equation}}
\newcommand{\eeq}{\end{equation}}

\newcommand{\ze}{\kern 0.05em}

\begin{document}

\title{ {\bf  Boson and Dirac stars in $D\geq 4$ dimensions}}
 \vspace{1.5truecm}
\author{
{\bf Jose Luis Bl\'azquez-Salcedo$^1$},
%\\
{\bf Christian Knoll$^1$} and
{\bf Eugen Radu$^{2}$}
\\
\\
% \vspace{0.5truecm}
%
{\small $^1$ Institut f\"ur  Physik, Universit\"at Oldenburg, Postfach 2503,
D-26111 Oldenburg, Germany}
\\
{\small 
$^2$
Departamento de F\'\i sica da Universidade de Aveiro and}
\\
{\small 
 Center for Research and Development in Mathematics and Applications (CIDMA),
}
\\
{\small 
 Campus de Santiago, 3810-183 Aveiro, Portugal
} 
}
%\date{August 2017}
\maketitle

\begin{abstract}   
We present a comparative study of spherically symmetric, 
localized,
particle-like solutions for spin $s=0,1/2$ and $1$ gravitating  fields
 in a $D$-dimensional, 
asymptotically flat spacetime.
These fields are massive, possessing a harmonic time dependence and no self-interaction.
Special attention is paid to
the  mathematical  similarities and  physical differences between the bosonic and fermonic cases.
We find that the generic pattern of solutions is similar for any
value of the spin $s$,
 depending only on the dimensionality of spacetime, the cases
$D=4,5$ being special.
 
\end{abstract}

%%%%%%%%%%%%%%%%%%%%%%%%%%%%%%%%%%%%%%%%%%%%%%%%%%%%%%%%%%
\section{Introduction and motivation}
%%%%%%%%%%%%%%%%%%%%%%%%%%%%%%%%%%%%%%%%%%%%%%%%%%%%%%%%%% 

The first explicit realization
of the idea of stable, localized bundles of energy as a model for particles
 can be traced back to the work of Kaup 
\cite{Kaup:1968zz}
and Ruffini and Bonazzalo 
\cite{Ruffini:1969qy},  
fifty years ago.
 They found asymptotically flat, spherically symmetric equilibrium solutions of the
Einstein-scalar field system in four spacetime dimensions.
These {\it boson stars} are macroscopic quantum states and are only prevented
from collapsing gravitationally by the Heisenberg uncertainty principle
\cite{Schunck:2003kk},
\cite{Liebling:2012fv}.
However, as shown in the recent work \cite{Brito:2015pxa},
 similar configurations exist also  
for a model with a gravitating massive spin-one field, representing {\it Proca stars}. 
Analogous solutions (although less known) were found also in
 Einstein-Dirac theory  \cite{Finster:1998ws},
the fermions being treated as  classical fields. 

A comparative study of all these cases can be found in  
the recent work
\cite{Herdeiro:2017fhv},
where it has been noticed that, 
as classical field theory solutions, 
the
existence of such self-gravitating  energy lumps does not distinguish
between the fermionic/bosonic nature of the fields, possessing a variety of similar features.
For example, in all  cases
there is a
harmonic
time dependence in the fields (with a frequency $w$),
together
with a confining mechanism,
as provided by a
 mass $\mu \neq 0$ of the elementary quanta of the field.
Moreover, 
when ignoring the
Pauli's exclusion principle,
the (field frequency-ADM mass)-diagram 
of the solutions
looks similar for both bosonic  and fermionic stars. 
Also, the existence of these (nontopological)  solitons can
be related to the fact that they possess a
 Noether charge $Q$. associated with a global U(1) global
symmetry.
 
As with other models, it is of interest to see how the dimensionality $D$ 
of spacetime affects the properties
of this type of solutions.
For example, 
one would like to know which of their properties
 are peculiar to four-dimensions, and which hold more generally. 
However, 
relatively little is known about the properties of
this type of configurations in $D>4$.
While $D=5$ spin-zero boson stars are discussed in \cite{Hartmann:2010pm},
the case of spin $s=1/2,1$ fields was not considered at all.
	
	\medskip
	
	The purpose of this work is to consider a comparative study of  
	scalar, Proca and Dirac particle-like solutions 
	in $D\geq 4$ dimensions, looking for spherically symmetric solutions which approach at infinity
	 a Minkowski spacetime background.
Our results show that,
despite the existence of some common features, the cases $D=4,5$	are rather special.
Perhaps the most striking feature is that 
the  $D>4$ solutions  configurations do not connect continuously to Minkowski spacetime vacuum, 
with the existence of a mass (and Noether charge) gap. 
Also, when ignoring the Pauli's exclusion principle,
the Dirac stars share the pattern of the bosonic configurations.

The paper is structured as follows:  in Section 2 we present the general  
framework for both geometry and matter fields.
Special attention is paid to the construction of a suitable 
spin $s=1/2$ field
Ansatz compatible with a spherically symmetric spacetime metric,
a task which, to our knowledge, has not been yet addressed in the literature.
In Section 3 the field equations are solved numerically, and
it is shown how the dimensionality of spacetime affects the properties of the solutions. 
We conclude with Section 4 where the results are compiled.

%%%%%%%%%%%%%%%%%%%%%%%%%%%%%%%%%%%%%%%%%%%%%%%%%%%%%%%%%%
\section{The framework}
%%%%%%%%%%%%%%%%%%%%%%%%%%%%%%%%%%%%%%%%%%%%%%%%%%%%%%%%%% 

%%%%%%%%%%%%%%%%%%%%%%%%%%%%%%%%%%%%%%%%%%%%%%%%%%%%%%%%%%
\subsection{The general action and line element}
%%%%%%%%%%%%%%%%%%%%%%%%%%%%%%%%%%%%%%%%%%%%%%%%%%%%%%%%%% 
We consider Einstein's gravity in $D$-spacetime dimensions 
minimally coupled with a spin-$s$ field  
(with $s=0,\frac{1}{2},1$), generically denoted as $\mathcal{U}$,
the corresponding action being
\begin{eqnarray}
\label{action}
\mathcal{S}=\int d^D x \sqrt{-g} 
\left [
\frac{1}{16 \pi G} R
+
\mathcal{L}_{(s)} 
\right] \ .
\end{eqnarray}
Extremizing the action (\ref{action}) leads to a system of coupled Einstein-matter equations of motion 
\begin{eqnarray}
 R_{\alpha \beta}-\frac{1}{2} R g_{\alpha \beta}=8 \pi G~T_{\alpha \beta}^{(s)},~~{\rm with}~~
T_{\alpha \beta}^{(s)}=\frac{2}{\sqrt{-g}}\frac{\delta \mathcal{L}_{(s)}  }{\delta g^{\alpha \beta}},
\end{eqnarray}
while the variation of  (\ref{action})  $w.r.t.$ $\mathcal{U}$ leads to the matter fields equations.

We are interested in horizonless,
nonsingular solutions describing particle-like, localised configurations with finite energy.
Restricting for simplicity to spherically symmetric configurations, 
the corresponding spacetime metric is most conveniently studied in Schwarzschild-like coordinates,
with
\begin{eqnarray}
\label{metric}
 ds^2=\frac{dr^2}{N(r)}+r^2 d\Omega_{D-2}^2-N(r)\sigma^2(r) dt^2\ ,
~~~{\rm with}~~N(r)\equiv 1-\frac{2m(r)}{r^{D-3}}\ ,
\end{eqnarray}
where  $d\Omega_{D-2}^2$ denotes
the line element on a $(D-2)-$dimensional sphere,
while $r$ and $t$ are the radial and time coordinates, respectively.
This Ansatz introduces two functions  $m(r)$ and $\sigma(r)$, with
$m(r)$ being related to the local mass-energy density up to some $D$-dependent factor.
Also, its asymptotic value, $m_0$,  fixes
 the total ADM mass of the spacetime,
\begin{eqnarray}
\label{masss}
 M= \frac{(D-2)V_{D-2}}{8\pi G}   m_0,
\end{eqnarray}
(with $V_{D-2}$ the area of the $(D-2)$-sphere).
An advantage of the above metric form  
is that it leads to simple first order equations for the functions $m, \sigma$,
\begin{eqnarray}
\label{eqE}
 m'=-\frac{8\pi G}{D-2}r^{D-2}T_t^t,~~\sigma'=\frac{8\pi G}{D-2}\frac{r \sigma}{N}(T_r^r-T_t^t).
\end{eqnarray}
The ground state of the model is  $\mathcal{U}=0$,
together with a flat spacetime metric ($i.e.$ $m=0$, $\sigma=1$).

%%%%%%%%%%%%%%%%%%%%%%%%%%%%%%%%%%%%%%%%%%%%%%%%%%%%%%%%%%
\subsection{The matter content}
%%%%%%%%%%%%%%%%%%%%%%%%%%%%%%%%%%%%%%%%%%%%%%%%%%%%%%%%%% 

In all three cases, the Lagrangian $\mathcal{L}_{(s)} $ possesses a
  $global$ $U(1)$ invariance, under the transformation $\mathcal{U} \rightarrow e^{i a}\mathcal{U} $, with $a$ being constant.
This	implies the existence of a conserved 4-current, $j^{\alpha}_{ (s) ;\alpha}=0$. 
Integrating the timelike component of this current on a spacelike slice yields a conserved quantity --
 the \textit{Noether charge}:
\begin{eqnarray}
\label{Q}
Q_{(s)}=V_{D-2}\int_{0}^\infty dr~ r^{D-2} \sigma~j^t _{(s)}\ .
\end{eqnarray}
 Upon quantization, $Q$ becomes an integer--the particle number.
Also, one remarks that 
the ADM mass $M$ and the Noether charge $Q$ provide the only global charges
of the system.

%%%%%%%%%%%%%%%%%%%%%%%%%%%%%%%%%%%%%%%%%%%%%%%%%%%%%%%%%%%%%%%%%%%%%%
\subsubsection{$s=0$: a massive complex scalar field}
%%%%%%%%%%%%%%%%%%%%%%%%%%%%%%%%%%%%%%%%%%%%%%%%%%%%%%%%%%%%%%%%%%%%%% 
We start with the simplest case of a 
 a complex scalar field $\Phi$ with a Lagrangian density
\begin{eqnarray}
\label{Ls}
 \mathcal{L}_{(0)}= - g^{\alpha \beta}\bar \Phi_{, \, \alpha} \Phi_{, \, \beta} - \mu^2 \bar \Phi \Phi,
\end{eqnarray}
the energy-momentum tensor, the current and the Klein-Gordon equation  being
\begin{eqnarray}
\label{Ts} 
&&
T_{\alpha \beta}^{(0)}
=
\bar  \Phi_{ , \alpha}\Phi_{,\beta}
+\bar \Phi_{,\beta}\Phi_{,\alpha} 
-g_{\alpha \beta}  \left[ \frac{1}{2} g^{\gamma \delta} 
 ( \bar \Phi_{,\gamma}\Phi_{,\delta}+
\bar \Phi_{,\delta}\Phi_{,\gamma} )+\mu^2 \bar \Phi\Phi\right],
\\
&&
j^\alpha_{(0)}=-i (\bar \Phi \partial^\alpha \Phi-\Phi \partial^\alpha \bar \Phi),
~~~ \nabla^2 \Phi-\mu^2\Phi=0.~{~}
\end{eqnarray}
A scalar field Ansatz which is compatible with a spherically symmetric 
geometry is written in terms of a single real function $\phi(r)$, and reads:
\begin{eqnarray}
\label{S}
&& 
\Phi=\phi(r)e^{-iw t} \ .
\end{eqnarray}
The scalar field amplitude solves the equation
\begin{eqnarray}
\label{eq-0}
\phi''+
\left(
\frac{D-2}{r}
+\frac{N'}{N}
+\frac{\sigma'}{\sigma}
\right)\phi'
+
(\frac{w^2}{N\sigma^2}-\mu^2)\frac{\phi}{N}=0,
\end{eqnarray}
while the Einstein equations imply
\begin{eqnarray}
\label{eq-0E}
 m'=\frac{8 \pi G}{D-2} r^{D-2}
\left(
N \phi'^2+\mu^2 \phi^2+\frac{w^2 \phi^2}{N\sigma^2}
\right),~~
\sigma'= \frac{16 \pi G}{D-2} r  \sigma
\left( 
\phi'^2+\frac{w^2\phi^2}{N^2\sigma^2}
\right).
\end{eqnarray}

%%%%%%%%%%%%%%%%%
Finally, following  \cite{HS,Heusler:1996ft},
one can prove the virial identity 
\begin{eqnarray}
	\int_0^\infty dr~r^{D-2} \sigma 
	\left(
	(D-3)\phi'^2+(D-1)\mu^2 \phi^2
	\right)
	=
	w^2 \int_0^\infty dr~ 
\left	(
2(D-2) N-D+3
\right) \frac{r^{D-2} \phi^2}{N^2\sigma},
\end{eqnarray}
 which shows $e.g.$ that a nontrivial scalar field
can only be supported if $w\neq 0$.

%%%%%%%%%%%%%%%%%%%%%%%%%%%%%%%%%%%%%%%%%%%%%%%%%%%%%%%%%%%%%%%%%%%%%%
\subsubsection{$s=1$: a massive complex vector field}
%%%%%%%%%%%%%%%%%%%%%%%%%%%%%%%%%%%%%%%%%%%%%%%%%%%%%%%%%%%%%%%%%%%%%% 
For any value of $D$,
the  complex Proca field is described by
a potential 1-form  $\mathcal{A}$ with the associated field strength $\mathcal{F}=d\mathcal{A}$
(where we denote the corresponding complex conjugates by an overbar, $\bar{\mathcal{A}}$ and $\bar{\mathcal{F}}$).
The corresponding Lagrangian density,  field equations, current and energy-momentum tensor are
\begin{eqnarray}
\label{LP}
&&
 \mathcal{L}_{(1)}= -\frac{1}{4}\mathcal{F}_{\alpha\beta}\bar{\mathcal{F}}^{\alpha\beta}
-\frac{1}{2}\mu^2\mathcal{A}_\alpha\bar{\mathcal{A}}^\alpha ,~
\nabla_\alpha\mathcal{F}^{\alpha\beta}-\mu^2 \mathcal{A}^\beta=0  ,~
j^\alpha_{(1)}=
\frac{i}{2}\left[\bar{\mathcal{F}}^{\alpha\beta}\mathcal{A}_\beta-\mathcal{F}^{\alpha \beta}\bar{\mathcal{A}}_\beta\right],
%\end{eqnarray}
%\begin{eqnarray}
\\
&&
T_{\alpha\beta}^{(1)}=\frac{1}{2}
( \mathcal{F}_{\alpha \sigma }\bar{\mathcal{F}}_{\beta \gamma}
+\bar{\mathcal{F}}_{\alpha \sigma } \mathcal{F}_{\beta \gamma}
)g^{\sigma \gamma}
-\frac{1}{4}g_{\alpha\beta}\mathcal{F}_{\sigma\tau}\bar{\mathcal{F}}^{\sigma\tau}+\frac{1}{2}\mu^2\left[  
\mathcal{A}_{\alpha}\bar{\mathcal{A}}_{\beta}
+\bar{\mathcal{A}}_{\alpha}\mathcal{A}_{\beta}
-g_{\alpha\beta} \mathcal{A}_\sigma\bar{\mathcal{A}}^\sigma\right] .~~~{~~~}
\end{eqnarray}
Note that the field equations imply the Lorentz condition (which for a Proca field is not a gauge choice, but a dynamical requirement),
%\begin{equation}
$\nabla_\alpha\mathcal{A}^\alpha= 0. $
%\label{lorentz}
%\end{equation}

The 1-form Ansatz compatible with a static, spherically symmetric geometry contains two real potentials, $F(r)$ and $G(r)$:
\begin{eqnarray} 
\mathcal{A}=\big(   F(r)dt+iG(r)dr    \big) e^{-iwt} ,
\label{Proca_ansatz}
\end{eqnarray}
which solve  the equations
\begin{eqnarray}
\label{eq-1}
F'=(w^2-\mu^2 N\sigma^2)\frac{G}{w},
~~
G'+
\left(
\frac{(D-2)}{r}+\frac{N'}{N}+\frac{\sigma'}{\sigma}
+\frac{w F }{N^2\sigma^2} \right)G=0,
\end{eqnarray}
 the corresponding equations for the metric functions being
\begin{eqnarray} 
\label{eq-1E}
m'= \frac{4\pi G}{D-2} r^{D-2}\left[\frac{(F'-wG)^2}{2\sigma^2}+ \mu^2\left(G^2N+\frac{F^2}{N\sigma^2}\right)\right],~~~
\sigma' = \frac{8\pi G}{D-2} \sigma r  \mu^2 \left(G^2+\frac{F^2}{N^2\sigma^2} \right).
\end{eqnarray}
%%%%%%%%%%%%%%%%%%%%%%%%%%%%%%%%%%%%%%%%%%%%%%%%%%%%%%%%%%%%%%%%%%%%%%
%
In this case the solutions satisfy the following  virial identity
\begin{eqnarray}
	\mu^2 \int_0^\infty dr~r^{D-2} \sigma 
	\left(
	(D-3)G^2-\frac{F^2(2(D-2)N-D+3)}{N^2\sigma^2}
	\right)
	= (D-3)\int_0^\infty dr~r^{D-2}  \frac{(wG-F')^2}{\sigma}.
\end{eqnarray}

%%%%%%%%%%%%%%%%%

\subsubsection{$s=1/2$: massive Dirac fields}
%%%%%%%%%%%%%%%%%%%%%%%%%%%%%%%%%%%%%%%%%%%%%%%%%%%%%%%%%%%%%%%%%%%%%% 
The case of a fermionic matter content is more subtle, since 
a model with a  single (backreacting)  spinor  
is not 
compatible with a spherically symmetric spacetime.
Thus, as in the $D=4$ case  \cite{Finster:1998ws}, 
one should consider several 
spinors with equal mass $\mu$ and the same frequency $w$,
each one possessing a specific angular dependence.
Although an individual energy-momentum tensor is not spherically symmetric,
the sum of all contributions
leads to a result which is compatible with the line-element (\ref{metric}).
In $D$-dimensions, such configuration can be constructed with
(at least)
 $n_f=2^{\left\lfloor \frac{D-2}{2} \right\rfloor}$
spinors $\Psi^{[A]}$ ($A=1...n_f$), 
each one with a Lagrangian density
\begin{eqnarray}
\label{LD} 
 \mathcal{L}_{(1/2)}^{[A]}=-i 
\left[
\frac{1}{2}
  \left( \{ \hat{\slashed D}  \overline{\Psi}^{[A]}  \} \Psi^{[A]} -
     \overline{\Psi}^{[A]} \hat{\slashed D}  \Psi^{[A]}
	\right)
+\mu \overline{\Psi}^{[A]}  \Psi^{[A]} 
\right],
%\end{eqnarray}
%%
%\begin{eqnarray}
\end{eqnarray}
the total energy-momentum tensor and the individual current being
\begin{eqnarray}
T_{\alpha \beta}^{(1/2)}=\sum_{A=1}^{n_f}
 T_{\alpha \beta}^{[A]},~{\rm with}~T_{\alpha \beta}^{[A]} =-\frac{i}{2} 
\left[ 
    \overline{\Psi}^{[A]} \gamma_{(\alpha} \hat{D}_{\beta)} \Psi^{[A]} 
-  \left\{ \hat{D}_{(\alpha} \overline{\Psi}^{[A]} \right\} \, \gamma_{\beta)} \Psi^{[A]} 
\right] ,~
 j^{\alpha[A]}_{(1/2)}=\bar \Psi^{[A]} \gamma^\alpha \Psi^{[A]} .~{~~}
\end{eqnarray}
Also, each spinor  solves the Dirac equation
\begin{eqnarray}
 \hat{\slashed D}\Psi^{[A]}   - \mu \Psi^{[A]}   = 0 ~.
 \label{eq_Dirac}
\end{eqnarray}

The Dirac equation in a $D> 4$ spherically symmetric background
has been extensively studied in the literature, see $e.g.$  
Refs. \cite{Dong:2003xy}.
Thus here we shall only review the basic steps,
together with the special choice of the Ansatz which leads 
to a total energy-momentum tensor
$T_{\alpha \beta}^{(1/2)}$
 which is compatible with the static and spherically symmetric line-element (\ref{metric}).
To achieve this aim,	
we impose the spinors to satisfy a set of conditions, 
which can be summarized as follows\footnote{For $D=4$,
the contruction of a spin
$s=1/2$ Ansatz compatible with a spherically symmetric spacetime is discussed at length in 
Ref.
\cite{Finster:1998ws} 
(although for rather different conventions than in this work).}.

i) {\it The separable Ansatz.} The first step is to assume some simple but generic Ansatz for each field $\Psi^{[A]}$. 
Since each individual spinor satisfies equation (\ref{eq_Dirac}), 
we want to make use of the separability of the angular and the radial dependence of each field. 
Hence we assume that:
\begin{itemize}
	\item the radial dependence is the same for all individual spinors;
	\item the temporal dependence is of the form of a phase  
	(similar to the previously considered $s=0,1$ fields), 
	and also common to all individual spinors;
	\item the only difference between spinors is in the angular part.
\end{itemize} 
Such an Ansatz takes the form
\begin{eqnarray}
\Psi^{[A]} = \mathrm e^{-\mathrm i w t} \phi_\kappa(r) \otimes \Theta_{(\kappa; A)} \, ,
\label{Dirac_spinor_Ansatz}
\end{eqnarray}
where $\phi_\kappa(r)$ only depends on the radial coordinate, and $\Theta_{(\kappa; A)}$ on the angular coordinates.

ii) {\it Solutions of the angular part.}
 In the second step, we solve the angular part of the decomposition
(\ref{Dirac_spinor_Ansatz}). 
Since each individual spinor satisfies the 
equation (\ref{eq_Dirac}), we make use of the separability of the angular and the radial dependence of each field \cite{Frolov:2017kze}. 
Also, it is convenient to choose a parametrization of the $(D-2)$-sphere  with
\begin{eqnarray}
\mathrm d \Omega_{D-2}^2 = \mathrm d \theta_1^2 + \sin^2{\theta_1} \, \mathrm d \phi_1^2 + \cos^2{\theta_1} \, \mathrm d \Omega_{D-4}^2 \, .
\end{eqnarray}
For $D>5$, a similar decomposition of $\mathrm d \Omega_{D-4}^2$ can be done by introducing further $\phi_j$ and $\theta_j$ coordinates on the $(D-4)$-sphere. 
Such a tower can be constructed until all the angles of the sphere are parametrized 
(with $d \Omega_{0}^2=0$, $d \Omega_{1}^2=d\phi_{\left\lceil \frac{D-2}{2} \right\rceil}^2$, $\lceil x \rceil$ denoting the ceiling function).
The decomposition makes the $\left\lceil \frac{D-2}{2} \right\rceil$ 
commuting azimuthal Killing directions $\phi_j$ explicit, and we have 
$\partial_{\phi_j} \Theta_{(\kappa; A)} = \mathrm i m_j \Theta_{(\kappa; A)}$ ($m_j$ being a half-integer). 

One can show 
 that the angular part of the spinor $\Theta_{(\kappa; A)}$ 
is an eigenfunction of a Dirac angular operator ${\mathcal K}_{D-2}$, with  $\kappa$  the corresponding angular eigenvalue
 \cite{Dong:2003xy,new}. 
Working with the vielbein
$
\boldsymbol{\omega}^t = \sqrt{N} \sigma \mathbf d t ,
$
$
\boldsymbol{\omega}^r = \frac{1}{\sqrt{N}} \mathbf d r ,
$
$
\boldsymbol{\omega}^j = r \, \boldsymbol{\omega}^j_{D-2} 
$
(where $j = 1, ..., D-2$ and $\boldsymbol{\omega}^j_{D-2}$ is a vielbein for the $(D-2)$-sphere),
 a tower of angular operators in the lower dimensional spheres is found, with 
$[\mathcal K_{D-2}, \mathcal K_{D-4}] = 0$, $[\partial_{\phi_1}, \mathcal K_{D-2}] = 0$, etc. 

As a result, with this parametrization, the angular dependence of the spinors can be explicitly solved, 
and the general solution is given by a combination of hypergeometric functions\footnote{These angular solutions are the spinor monopole harmonics for a $(D-2)$-sphere \cite{Wu:1976ge}; 
also, they are related with the solutions of the angular part 
in \cite{Blazquez-Salcedo:2017bld}.}.
However, for our purpose, it is enough to consider the case where the angular eigenvalue is minimal 
(which means that each individual spinor carries the minimally allowed angular momentum value).
 The analysis of the solutions shows that this happens when 
\begin{eqnarray}
\label{spirality}
\kappa=\pm \frac{(D-2)}{2}~~{\rm and}~~~  m_1 = \pm \frac{1}{2} . 
\end{eqnarray}
Apart from the global sign of $\kappa$, which determines the ``$spirality$'' of the configuration,
 in practice, the solution on the $(D-2)$-sphere possesses
$\left\lfloor \frac{D-2}{2} \right\rfloor$ free sign combinations ($\lfloor x \rfloor$ denoting the floor function).

iii) {\it Combination of the angular part.}
Hence, the third step is to select a proper combination of all these different angular parts of the spinors, 
making use of all these free signs. 
One can easily prove that, if one chooses $n_f=2^{\left\lfloor \frac{D-2}{2} \right\rfloor}$ spinors, 
the angular parts can be combined in such a way that almost all the non-diagonal 
components of the total energy-momentum tensor are zero \cite{new}. 
As a result, the total combination of fields carries zero angular momentum. 
The only non-diagonal component of the total energy-momentum tensor that cannot be set 
to zero 
just by making use of the angular dependence of the spinors is the $(t,r)$-component. 

iv) {\it Vanishing radial current.} 
One remarks that  $T_r^t\neq 0$
is related to the existence of a non-vanishing radial current 
for the generic Ansatz (\ref{Dirac_spinor_Ansatz}). 
Nonetheless, in order to set it to zero it is enough to impose 
one condition on the radial part of the spinors Ansatz. 
For simplicity, it is convenient to fix the representation of the radial $\gamma$ matrices 
at this stage\footnote{The 
representation of the angular $\gamma$ matrices is that employed in
	\cite{Blazquez-Salcedo:2017bld}.},
\begin{eqnarray}
\gamma^t = \left[ \begin{array}{cc} 0 & 1 \\ 1 & 0 \end{array} \right] \; , \; \gamma^r = \left[ \begin{array}{cc} 0 & -1 \\ 1 & 0 \end{array} \right] \; , \; \phi_\kappa = \left[ \begin{array}{c} \phi_1 \\ \phi_2 \end{array} \right] \, ,
\end{eqnarray}
with $\phi_1$, $\phi_2$ two complex functions depending only on $r$.
Then the vanishing of the radial current, $j^{r[A]}_{(1/2)}=0$, for the Ansatz 
(\ref{Dirac_spinor_Ansatz})
implies that $|\phi_1|^2 = |\phi_2|^2$. 
One can choose
\begin{eqnarray}
\phi_1 = 
\hat \phi \; , \; 
\phi_2 = 
 \mathrm{e}^{\mathrm i \nu} \hat \phi \, ,
\end{eqnarray}
with $\hat \phi$ a complex function and $\nu$ a real valued function \footnote{However, let us comment here that in the field equations, the phase of $\hat \phi$ plays no role, and thus one can choose without loss of generality $\hat \phi$ to be a real function).}.
 This is enough to assure that the $(t,r)$-component of the $total$ energy-momentum tensor vanishes.

\medskip

With all these requirements, the total energy-momentum tensor is diagonal and compatible with the static and spherically symmetric line element (\ref{metric}).
 
To simplify the numerical treatment of the problem
it is convenient to define
\begin{eqnarray}
 \hat \phi \mathrm{e}^{\frac{i \nu}{2}}=
  g-if~,
\end{eqnarray}
the equations for the new matter functions being
\begin{eqnarray}
\label{deq_dirac}
f'+
\left(
\frac{N'}{N}+\frac{2\sigma'}{\sigma}+\frac{2(D-2)}{r} +\frac{2(D-2)}{r\sqrt{N}}
\right)\frac{1}{4}f
+
(
\frac{\mu}{\sqrt{N}}-\frac{w}{ N\sigma}
)g=0,
\\
g'+
\left(
\frac{N'}{N}+\frac{2\sigma'}{\sigma}+\frac{2(D-2)}{r} -\frac{2(D-2)}{r\sqrt{N}}
\right)\frac{1}{4}g
+
(
\frac{\mu}{\sqrt{N}}+\frac{w}{ N\sigma}
)f=0~.
\end{eqnarray}
Also,
the corresponding equations for the metric functions are
\begin{eqnarray}
\label{deq_metric}
 m'=\frac{64 \pi G r^{D-2} }{D-2} \frac{w(f^2+g^2)}{\sqrt{\sigma}N},
~~
\frac{\sigma'}{\sigma}= \frac{64 \pi G}{D-2}\frac{r}{N}
\left(
-\frac{(D-2)fg}{r}
+\frac{2w(f^2+g^2)}{\sigma \sqrt{N} }
+\mu(f^2+g^2)
\right).
\end{eqnarray}

%\textbf{
	Note that in the above equations (\ref{deq_dirac})-(\ref{deq_metric}) we have fixed the ``spirality'' to $+1$, 
the only case discussed in this work. 
	Nonetheless, all the qualitative properties of the Dirac stars that we present in the next Section are not affected by this choice
	 (for instance, other properties of the Dirac field on static and spherically symmetric configurations, such as the quasinormal modes, are not qualitatively affected by the change of spirality \cite{Blazquez-Salcedo:2017bld}).
 %}

%%%%%%%%%%%%%%%%%%%%%%%%%%%%%%%%%%%%%%%%%%%%%%%%%%%%%%%%%%%%%%%%%%%
\begin{figure}[h!]
\begin{center}
\includegraphics[width=0.45\textwidth]{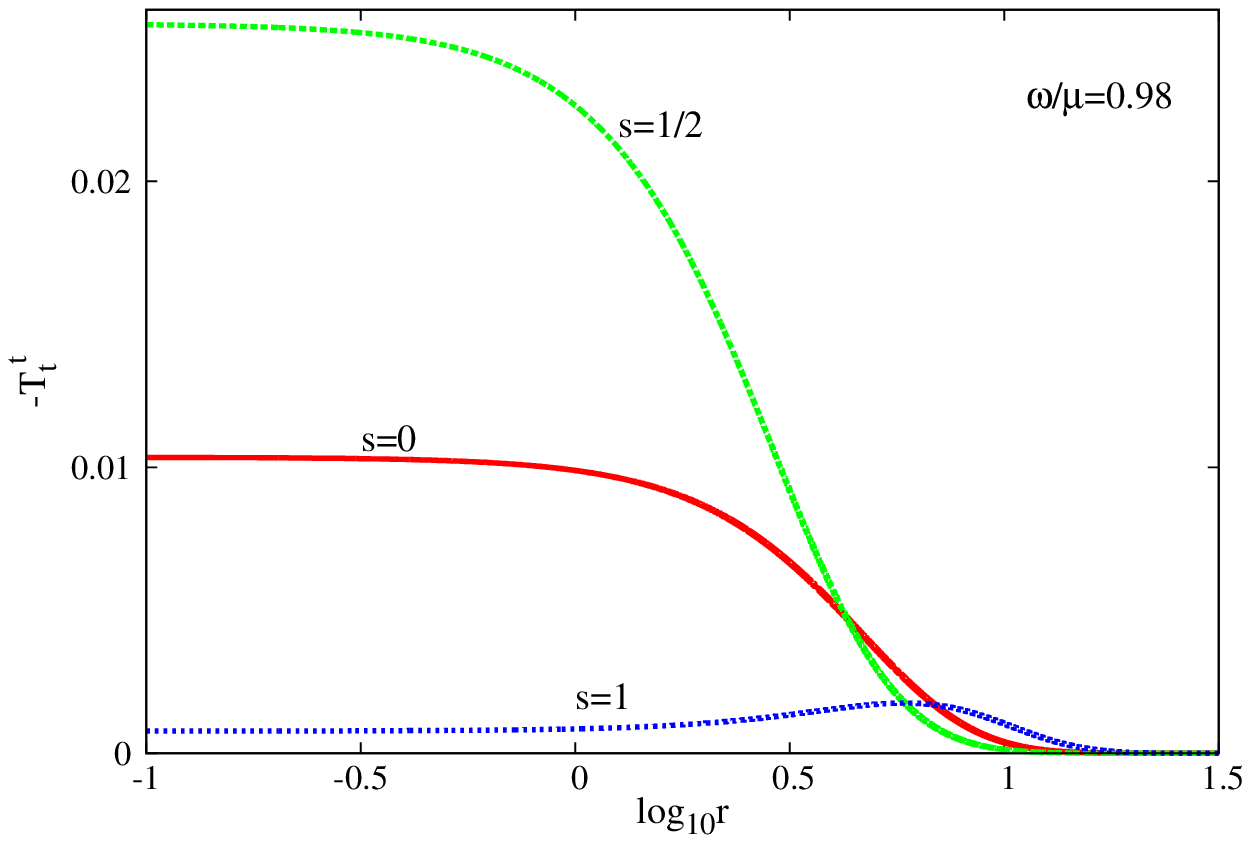}\ \ \
\includegraphics[width=0.45\textwidth]{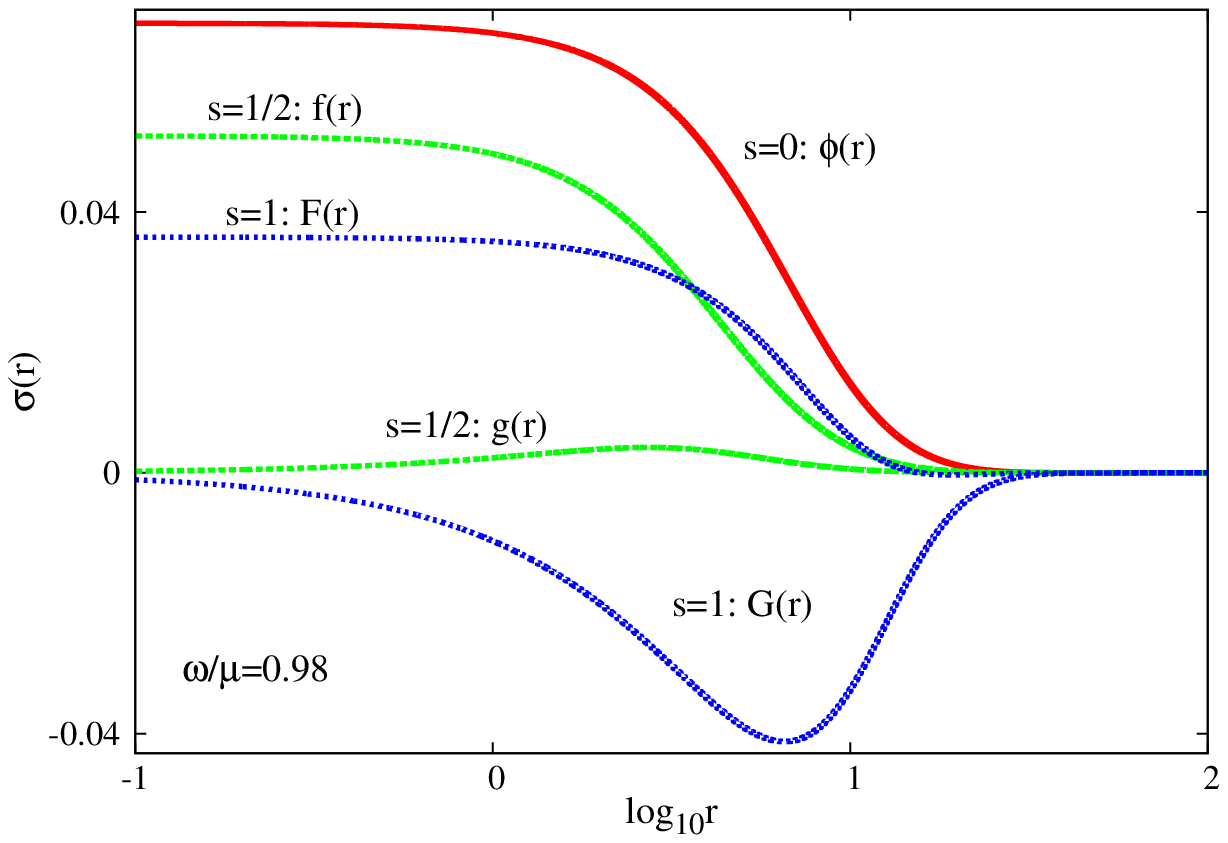}
\includegraphics[width=0.45\textwidth]{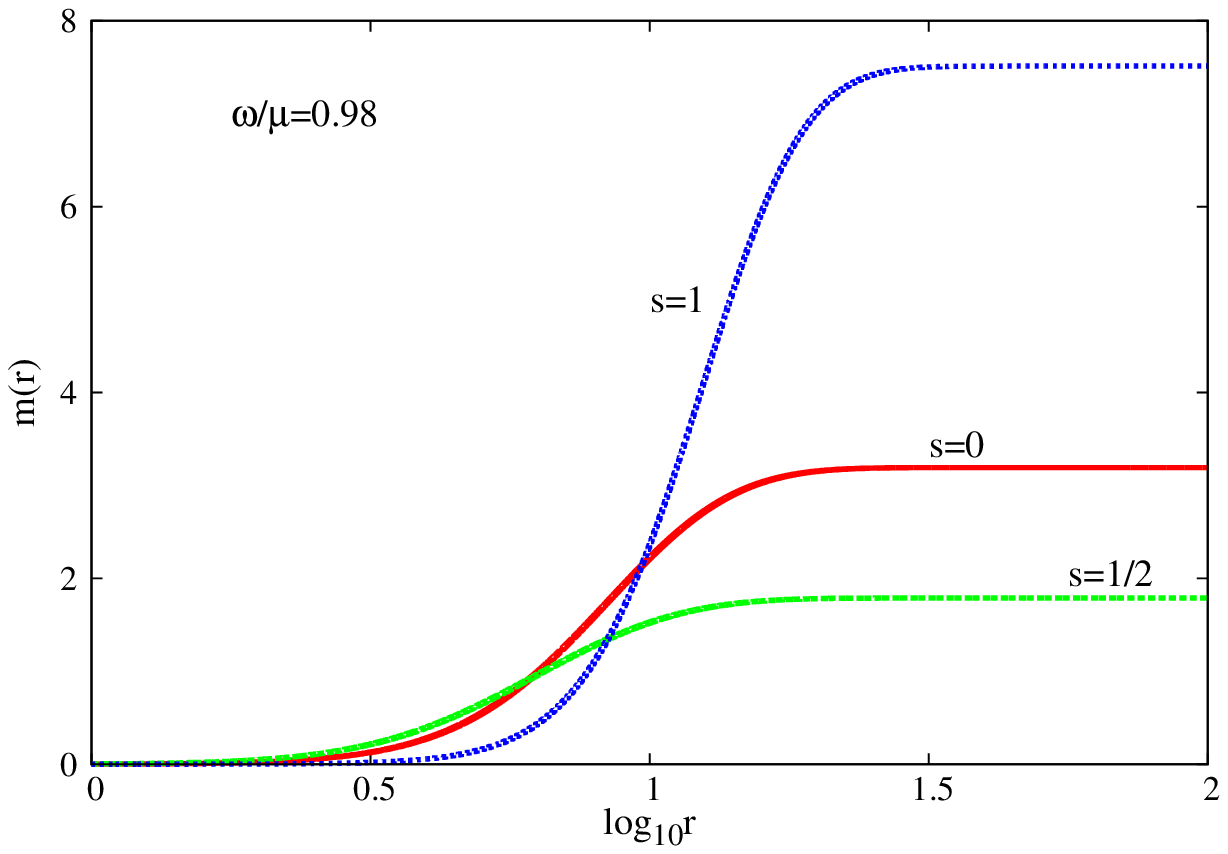}\ 
\includegraphics[width=0.45\textwidth]{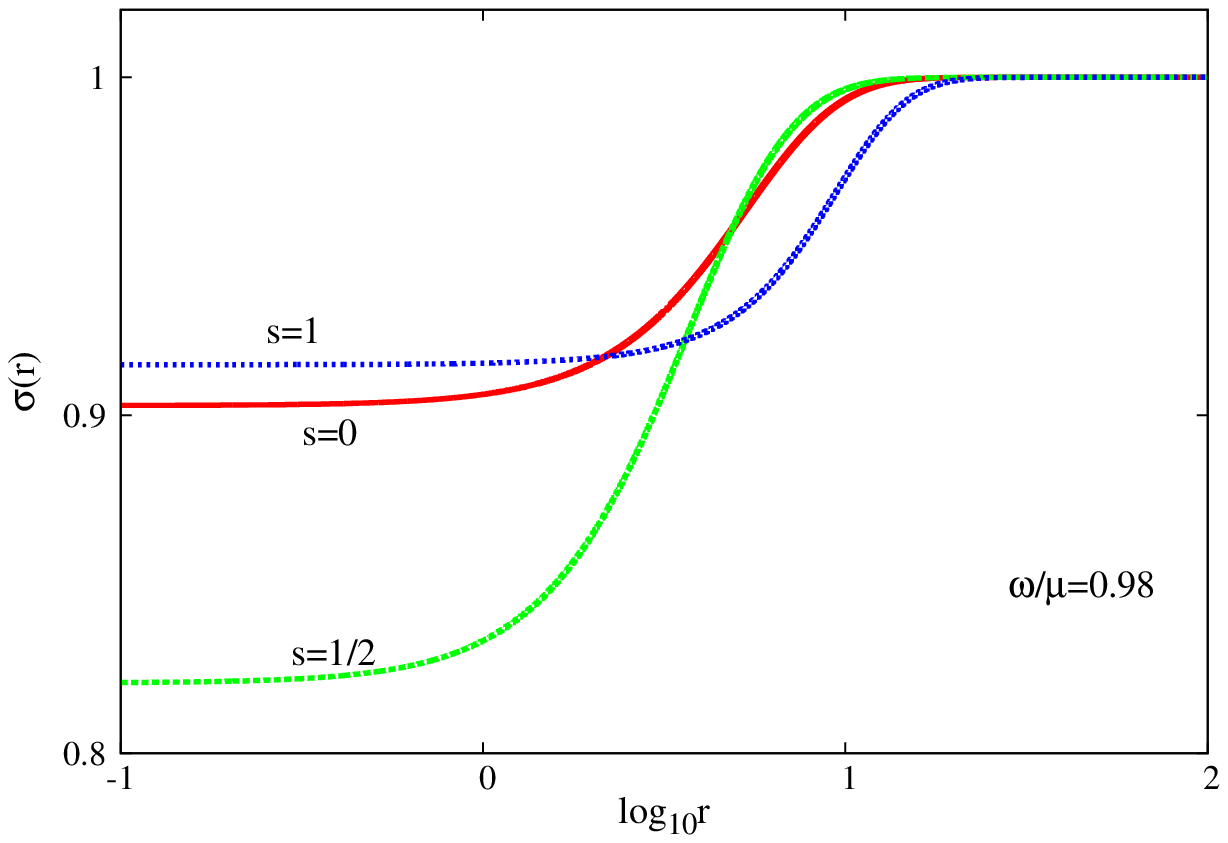}
\caption{\small{{\it Top left panel:} the energy density  is shown for a (typical) solution of each 
spin-$s$ model in $D=5$ dimensions, 
all with the same frequency to particle mass ratio, $w/\mu=0.98$. 
{\it Top right panel:} the matter functions profiles for the same solutions.
 {\it Bottom panels:} the metric functions are shown for the same solutions. }}
\label{profile}
\end{center}
\end{figure}  
%%%%%%%%%%%%%%%%%%%%%%%%%%%%%%%%%%%%%%%%%%%%%%%%%%%%%%%%%%%%%%%%%%%

Finally, the virial identity satisfied by the solutions reads
\begin{eqnarray}
	&&
	\int_0^\infty dr~r^{D-2} \sigma 
	\left(
	(D-2)^2\frac{fg}{r}+\frac{(2(D-1)+(D+1)(N-1))}{2\sqrt{N}}(gf'-fg')
	\right)
	=
	\\
	&&
	\nonumber
	\int_0^\infty dr~r^{D-2} \sigma 
	\left(
	\frac{w(f^2+g^2)}{2N\sqrt{N}}(2(D-1)+(3D-5)(N-1))
	+(D-1)\mu(f^2-g^2)
	\right).
\end{eqnarray}

%%%%%%%%%%%%%%

%%%%%%%%%%%%%%%%%%%%%%%%%%%%%%%%%%%%%%%%%%%%%%%%%%%%%%%%%%%%%%%%%%%%%%
\section{The results}
%%%%%%%%%%%%%%%%%%%%%%%%%%%%%%%%%%%%%%%%%%%%%%%%%%%%%%%%%%%%%%%%%%%%%% 

We are interested in particle-like  solutions
of the eqs.
 (\ref{eq-0}), (\ref{eq-0E}), 
(\ref{eq-1}), (\ref{eq-1E})
and
(\ref{deq_dirac}),
(\ref{deq_metric}),
with a topologically trivial, smooth geometry and a regular matter distribution.
Thus,
as $r\to 0$, one imposes $N(r) \to 1$ (with $m(r)\sim O(r^{D-1})$), while $\sigma(r)\to \sigma_0>0$.
The requirement of finite mass and asymptotic flatness
imposes $m(r)\to m_0$ 
($i.e.$ $N(r)\to 1$) and $\sigma(r)\to 1$
as $r\to \infty$. 
Also, all matter functions vanish ar infinity,
%$
%
%$
while their behaviour near the origin is more complicated, with
$f(r)$ and $G(r)$
vanishing there,
while
$\phi(r)$,
$g(r)$
and 
$F(r)$
satisfy Neumann boundary conditions.
%$
%
For any $s$,
an approximate form of the solutions can be systematically constructed in both regions,
near the origin and for large values of $r$. 
For example, the near-origin expansion contains
two free parameters, one of them being $\sigma(0)$, and the other one being
$\phi(0)$, $g(0)$ or $F(0)$,
while the matter fields decay exponentially in the far field.

The solutions that smoothly interpolate between these asymptotics are constructed numerically.
The results reported in this work are found in units with $\mu=1$, $4\pi G=1$ 
(thus we use a scaled radial coordinate $r\to r/ \mu$ 
(together with $m\to m /\mu^{D-3}$, $w\to w  \mu$
and, for $s=1/2$, $f\to  f \sqrt{\mu}$, $g\to  g \sqrt{\mu}$)
while the factor of $4\pi G$ is absorbed in the expression of the matter functions).
The equations are solved by using a standard Runge-Kutta ODE solver 
and implementing a shooting method in terms of the near-origin essential parameters,
 integrating
towards $r\to \infty$.

For a given spin-$s$ model,
the only input parameters are the number $D$ of spacetime dimensions and the value $w$
of the frequency.
Then a (presumably infinite) set of solutions is found for some range of $w$, as indexed 
by the number of nodes of the matter functions.
Note that, however, only fundamental solutions are reported in this work.

We have constructed in a systematic way scalar, Proca and Dirac stars in $D=4,5$ and $6$ dimensions;
partial (less accurate) results were also found for $D=7$, $8$. 
The profile of typical $D=5$  configurations
with the same ratio $w/\mu$ are shown in
Fig. \ref{profile}, together with the corresponding 
energy-density distribution. 
This plot appears to be generic, a (qualitatively) similar picture being found for other
values  of $w$ (or even for $D\neq 5$).

%%%%%%%%%%%%%%%%%%%%%%%%%%%%%%%%%%%%%%%%%%%%%%%%%%%%%%%%%%%%%%%%%%%%%%%%%%%%%%%%%%%%%%
 \begin{figure}[h!]
\begin{center}
\includegraphics[width=0.495\textwidth]{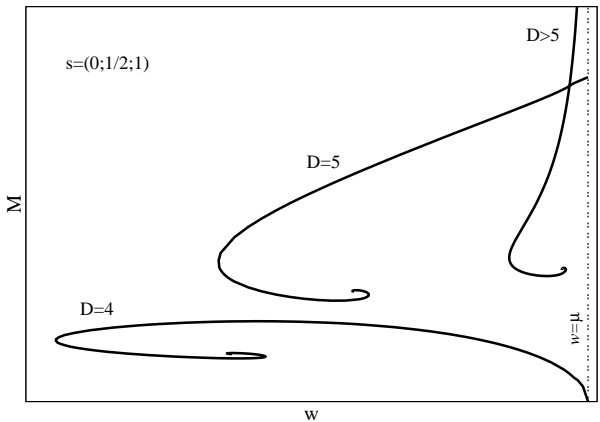} 
\caption{\small{ The generic $M(w)$-diagram (with $M$ the ADM mass and $w$ the field frequency) 
is shown for the $s=0,1/2,1$ families of solutions in $D\geq 4$ dimensions.
}}
\label{fig-generic}
\end{center}
\end{figure} 
%%%%%%%%%%%%%%%%%%%%%%%%%%%%%%%%%%%%%%%%%%%%%%%%%%%%%%%%%%%%%%%%%%%%%%%%%%%%%%%%%%%%%% 

\begin{figure}[h!]
\begin{center}
\includegraphics[width=0.49\textwidth]{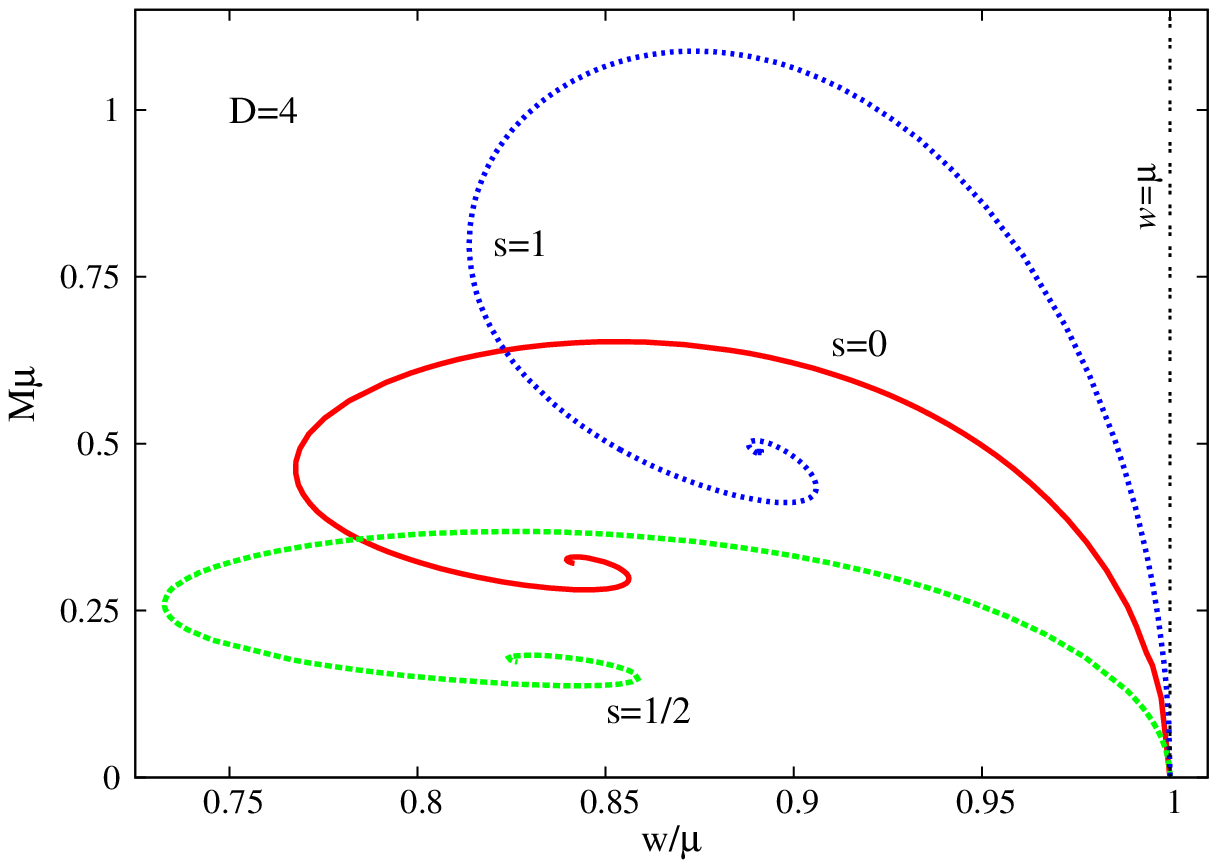}\ \ \
\includegraphics[width=0.49\textwidth]{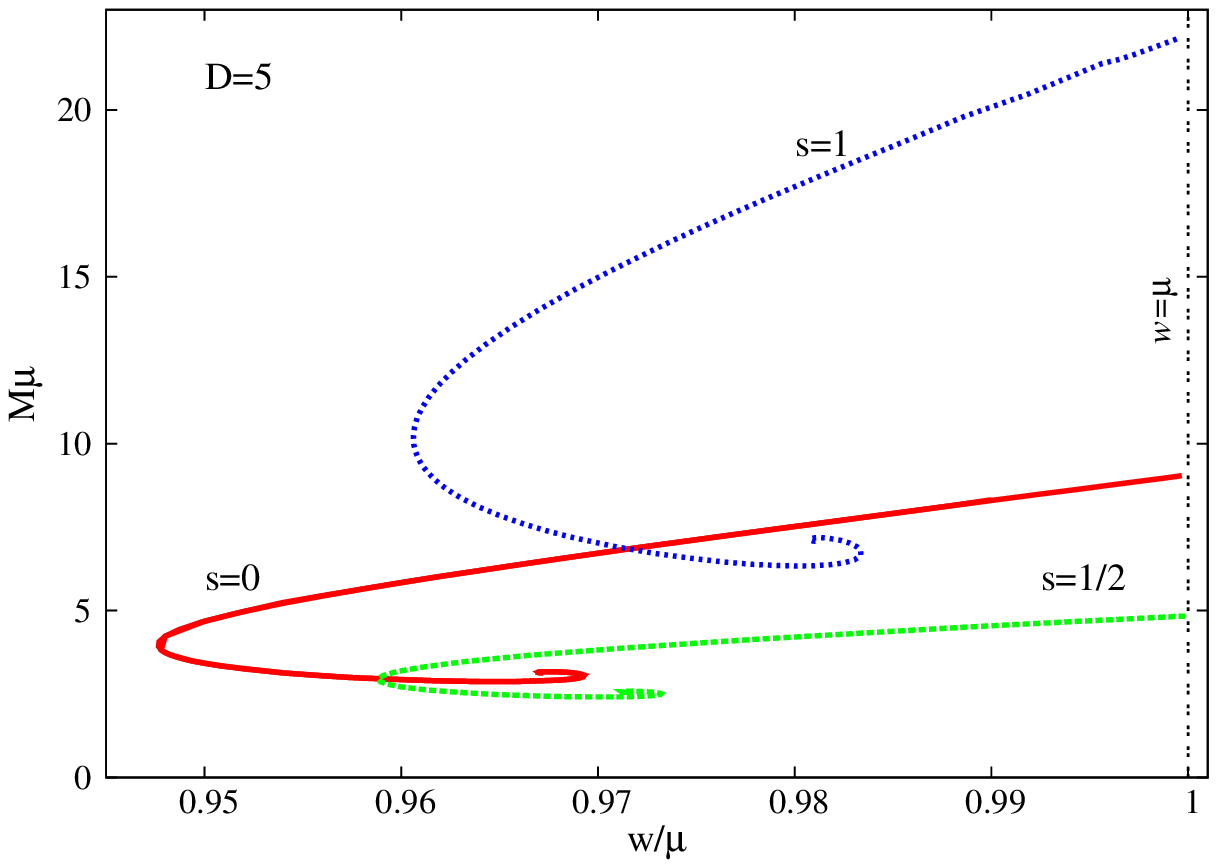}
\includegraphics[width=0.49\textwidth]{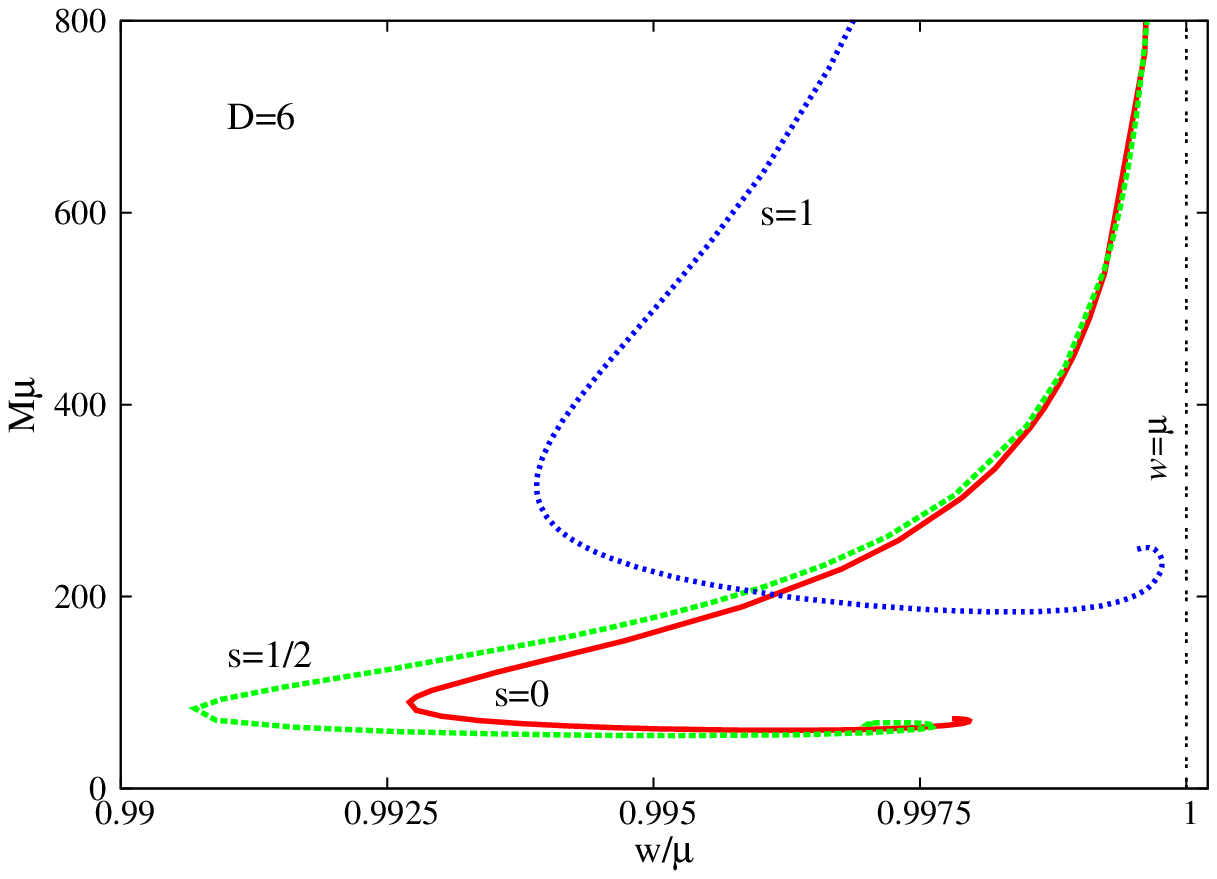}
\caption{\small{The ADM mass $M$  $vs.$ field frequency $w$ is shown for scalar ($s=0$),
Proca ($s=1$} and Dirac $(s=1/2)$ stars in $D=4,5,6$ dimensions.}
\label{results}
\end{center}
\end{figure}

Remarkably, the $s=0,1/2,1$ solutions exhibit a certain degree of universality,
the pattern depending on the number of the spacetime dimensions.
Their basic (qualitative) features are displayed in Fig. \ref{fig-generic} in a frequency-mass diagram,
which is the main result of this work,
while quantitative results are shown in Fig.  \ref{results}.  

The study of these plots indicates the existence of a number of 
basic properties which hold for both bosonic and fermionic solutions  and can be summarized as follows.
\begin{itemize}
\item
For any $D$, a (continuous) family of $s=0,1/2,1$ solutions exists for a limited range of frequencies only,
$w_{min}<w<\mu$,
the minimal value of the ratio $w/\mu$ decreasing with the spacetime dimension.
After reaching the minimal frequency, the $M(w)$-curve backbends into a second branch; 
moreover, 
further branches  and backbendings are found.
We conjecture  that, for any $D$, the $M(w)$-curve becomes a spiral,  which 
  approaches  at its center  a critical solution,
a result rigorously established  so far only for the $D=4,~s=0$ case
\cite{Friedberg:1976me},
\cite{Friedberg:1986tp}. 
\item
The behaviour of the solutions as $w\to \mu$ depends on the number of spacetime dimensions.
For $D=4$, the matter field(s) becomes very diluted and the solutions trivialize in that limit, with $M\to 0$,
the maximal mass value being attained
at some intermediate frequency.

The picture for $D=5$ is different and 
a mass gap is found   between the  $\mathcal{U}=0$ 
vacuum Minkowski ground state and the
set of solutions with $\mathcal{U} \neq 0$.
Moreover, the
 limiting configurations with a frequency $w$ arbitrarily close to $\mu$ exhibit a different pattern
as compared to the four dimensional case.
That is, although the matter fields spread  and tend to zero, while the geometry
becomes arbitrarily close to that of flat spacetime,  the mass remains
finite and nonzero (being also the maximal allowed value).

The existence of a  mass gap is a property found also for
 $D\geq 6$ solutions.
Also, again
 the solutions do not trivialize as  $w \to \mu$. 
However, in this case their mass appears to diverge in that limit.
\item
A similar picture is found when considering instead a $Q(w)$
diagram.
In particular, one finds that $M>\mu Q$ for all $D>4$ solutions, with (at the level of numerical accuracy)
$M\to \mu Q$ as $w\to \mu$.
The picture for $D=4$
is more complicated, with $M<\mu Q$ for some range of frequencies ranging between $\mu$
and some critical value 
\cite{Schunck:2003kk},
\cite{Liebling:2012fv}.
\end{itemize}

%%%%%%%%%%%%%%%%%%%%%%%%%%%%%%%%%%%%%%%%%%%%%%%%%%%%%%%%%%%%%%%%%%%
%%%%%%%%%%%%%%%%%%%%%%%%%%%%%%%%%%%%%%%%%%%%%%%%%%%%%%%%%%%%%%%%%%%%%%

%%%%%%%%%%%%%%%%%%%%%%%%%%%%%%%%%%%%%%%%%%%%%%%%%%%%%%%%%%%%%%%%%%%%%%% 
 \begin{figure}[h!]
\begin{center}
\includegraphics[width=0.495\textwidth]{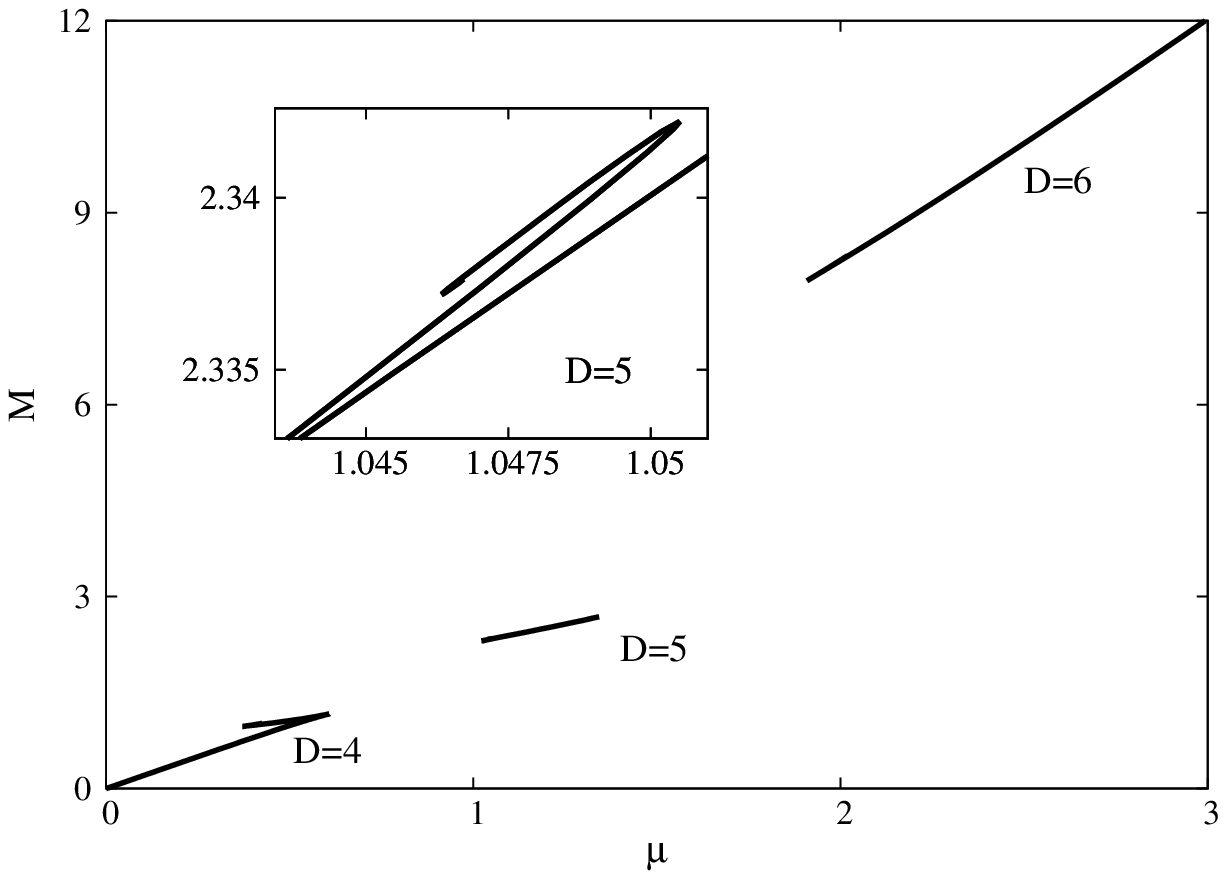} 
\caption{\small{ The ADM mass is shown vs. field mass $\mu$ for Dirac stars in  
$D=4,~5$ and $6$ dimensions.
 The single particle condition, $Q = 1$ for a fermionic field, is imposed here.
Note that the $D=6$ curve continues to large values, presumably diverging. The inset focuses on the $D=5$ case.
}}
\label{fig-Dirac}
\end{center}
\end{figure} 
%%%%%%%%%%%%%%%%%%%%%%%%%%%%%%%%%%%%%%%%%%%%%%%%%%%%%%%%%%%%%%%%%%%%%% 

%%%%%%%%%%%%%%%%%%%%%%%%%%%%%%%%%%%%%%%%%%%%%%%%%%%%%%%%%%%%%%%%%%%%%%
\subsection{Fermions: the single particle condition}
%%%%%%%%%%%%%%%%%%%%%%%%%%%%%%%%%%%%%%%%%%%%%%%%%%%%%%%%%%%%%%%%%%%%%% 

In the discussion above, no distinction has been made between the
bosonic and fermionic solutions. 
However,
although we have treated the Dirac equation
classically, 
the fermionic nature of a spin $s=1/2$ field would manifest at the level of the occupation number.
Thus, for each field $\Psi^{[A]} $,
 at most a single
particle should exist, in accordance to Pauli's exclusion principle.

The single particle condition, $Q=1$ is imposed by using the following scaling symmetry of the
Einstein-Dirac system\footnote{Note that $Q$ corresponds to the Noether charge for a single spinor,
the total charge being $Q_{tot}=\sum_A Q=2^{\left\lfloor \frac{D-2}{2} \right\rfloor}Q$.}
:
\begin{eqnarray}
\nonumber
r \to \lambda \bar r,~w\to \frac{\bar w}{\lambda},~\mu\to \frac{\bar \mu}{\lambda},~m\to \lambda^{D-3}\bar m,~
f\to \frac{\bar f}{\sqrt{\lambda}},~g\to \frac{\bar g}{\sqrt{\lambda}},~
{\rm while}~M\to  \lambda^{D-3}\bar M,~Q\to \lambda^{D-2}\bar Q,
\end{eqnarray}
(with $\lambda>0$ arbitrary and $\sigma$ invariant).
One can easily verify that this transformation does not affect the equations of motion;
however, it $changes$ the model, since it leads to a different field mass $\mu$. 
Then, for an initial solution with a given $Q$, 
the condition $\bar Q=1$ is imposed by taking  
$\lambda=Q^{\frac{1}{D-2}}$, 
which accordingly changes the corresponding values for the ADM mass and the field's frequency and mass. 

The resulting $M(\mu)$-diagram  is shown in Fig. \ref{fig-Dirac}
for Einstein-Dirac solutions in $D=4,~5$ and 6 dimensions 
(note that there we drop the overline for 
both $M$ and $\mu$).
Again, one notices a different behaviour in each spacetime dimension.
In four dimensions, the ($n_f=2$ particle) Einstein-Dirac solutions exist for 
a family of models with
both $\mu$ and $M$ ranging from zero to a maximal value of order one \cite{Herdeiro:2017fhv}.
The situation is different in $D=5$, in which case the minimal value of  $\mu$ and $M$
is nonzero (and $n_f=2$ again). 
This feature is preserved by $D=6$
solutions (with $n_f=4$ particles);
however,
no upper bounds for $\mu$ and $M$
appear to exist in that case.

{ In the inset of Fig. \ref{fig-Dirac}, we zoom a part of the $D=5$ curve, revealing a structure of peaks.
 This behaviour is generic for any dimension $D$, and it is related with the inspiral behaviour 
commented in the previous section. 
As a result, one can see that for fixed values the dimension $D$ and the scaled field mass $\mu$
($i.e.$ for fixed models), it is possible to have situations with a single solution (only one value of $M$), a discrete set of solutions (with different values of $M$), or no solution.}
 %

%%%%%%%%%%%%%%%%%%%%%%%%%%%%%%%%%%%%%%%%%%%%%%%%%%%%%%%%%%%%%%%%%%%%%%
\section{Further remarks}
%%%%%%%%%%%%%%%%%%%%%%%%%%%%%%%%%%%%%%%%%%%%%%%%%%%%%%%%%%%%%%%%%%%%%% 

The main purpose of this work was to provide a preliminary investigation of 
a special type of solitonic
solutions of gravitating
 matter systems in a number $D\geq 4$ of spacetime dimensions. 
The (massive) matter fields correspond  to bosons (spin 0, 1) or Dirac fermions (spin $1/2$), respectively,
and possess an harmonic time dependence.
The simplest solutions of this type are the well-known $D=4$, $s=0$ boson stars 
\cite{Kaup:1968zz},
\cite{Ruffini:1969qy}. 

Our results show that, for any $D$, the
existence of these  particle-like solutions, does not distinguish
between the fermionic/bosonic nature of the field, 
with the existence of a general pattern fixed by the number of spacetime dimensions.
Moreover,  while
the cases $D=4$ and $D=5$ are special,  
the $D>5$ solutions appear to share the same (qualitative) picture.
Perhaps the most curious feature revealed by this study is
the existence, for $D>4$, of a  mass gap,
the set of spin-$s$ configurations being not continuously connected to Minkowski spacetime vacuum.

\medskip
 
Among the numerous avenues that one may pursue following
the study here we mention a few. 
First, it would be interesting to clarify the stability of $D>4$ solutions,
an issue which has been extensively studied for $D=4$
(see $e.g.$
\cite{Schunck:2003kk},
\cite{Liebling:2012fv}
for $s=0$,
\cite{Brito:2015pxa},
\cite{Sanchis-Gual:2017bhw}
for $s=1$,
and
\cite{Finster:1998ws}
for the $s=1/2$ case).
Since, as noticed above, the higher dimensional configurations have $M>\mu Q$,
they possess an excess energy and  thus we expect them to be unstable against fission
(the  case $s=1/2$ being more subtle, due to the single particle condition).
A better understanding of the limiting behaviour of the solutions as $w\to \mu$
and towards the center of the $M(w)$-spiral
is another important open question.
In the scalar field case (with $D=4$),
an explanation of this behaviour is provided in
\cite{Friedberg:1976me},
\cite{Friedberg:1986tp}
(see also the explanation in
 Ref. \cite{Hartmann:2010pm} for the beaviour
of the $D=5$
bosons stars ($s=0$)
 as $w\to \mu$ ).

Furthermore, 
one may inquire if, similar to other field theory models \cite{Volkov:1998cc},
 these spin-$s$
solitons  possess generalizations with 
  a horizon at their center.
In four dimensions, no such (spherically symmetric) solutions exist,
as shown in \cite{Pena:1997cy} for spin 0,
 in 	\cite{Herdeiro:2016tmi} for spin 1
and in  
\cite{Finster:1998ak},
\cite{Finster:1998ju}
 for a Dirac field.
We expect a similar result to hold also in the higher dimensional case.

Finally, it would be interesting to investigate the solutions in this work
within the  framework of the large-$D$ 
limit of General Relativity \cite{Emparan:2013moa}.

\medskip
%
%%%%%%%%%%%%%%%%%%%%%%%%%%%  
\section*{Acknowledgements}
%%%%%%%%%%%%%%%%%%%%%%%%%%%
JLBS and CK would like to acknowledge support by the
DFG Research Training Group 1620 {\sl Models of Gravity}.
JLBS would like to acknowledge support from the DFG project BL 1553. 
The work of E.R. was supported by
 Funda\c{c}\~ao para a Ci\^encia e a Tecnologia (FCT),
within project UID/MAT/04106/2019 (CIDMA),
and by national funds (OE), through FCT, I.P., in the scope
of the framework contract foreseen in the numbers 4, 5 and 6
of the article 23, of the Decree-Law 57/2016, of August 29,
changed by Law 57/2017, of July 19.
E.R. also acknowledge support by the FCT grant PTDC/FIS-OUT/28407/2017.
This work has further been supported by  the  European  Union's  Horizon  2020 
 research  and  innovation  (RISE) programmes H2020-MSCA-RISE-2015
Grant No.~StronGrHEP-690904 and H2020-MSCA-RISE-2017 Grant No.~FunFiCO-777740. 
The authors would also like to acknowledge networking support by the
COST Action CA16104 {\sl GWverse}.

%%%%%%%%%%%%%%%%%%%%%%%%%%%%%%%%%%%%%%%%%%%%%%%%%%%%%%%%%%%%%%%%%%%%%%%%%%%%%%

%%%%%%%%%%%%%%%%%%%%%%%%%%%%%%%%%%%%%%%%%%%%%%%%%%%%%%%%%%%%%%%%%%%%%%%%%%%%%%

\end{document}